\documentclass{PoS}

\newcommand{\ba}{\begin{eqnarray*}}
\newcommand{\ea}{\end{eqnarray*}}
\newcommand{\be}{\begin{equation}}
\newcommand{\ee}{\end{equation}}
\newcommand{\bd}{\begin{displaymath}}
\newcommand{\ed}{\end{displaymath}}
\newcommand{\Eq}[1]{Eq.\,#1}

\newcommand{\Fig}[1]{Fig.\,#1}

\newcommand{\amu}{a_\mu}

\newcommand{\alhvp}{a_{l}^\mathrm{hvp}}
\newcommand{\aehvp}{a_{e}^\mathrm{hvp}}
\newcommand{\amuhvp}{a_{\mu}^\mathrm{hvp}}
\newcommand{\atauhvp}{a_{\tau}^\mathrm{hvp}}
\newcommand{\albarhvp}{a_{\overline{l}}^\mathrm{hvp}}

\newcommand{\amubarhvp}{a_{\overline{\mu}}^\mathrm{hvp}}

\title{Leading-order hadronic contribution to g-2 from lattice QCD}

\ShortTitle{Leading-order hadronic contribution to g-2 from lattice QCD}

\author{\speaker{Dru B.\ Renner}\thanks{Current address:\ Jefferson Laboratory.}\\
NIC, DESY, Platanenallee 6, D-15738 Zeuthen, Germany\\
E-mail:\ \email{dru@jlab.org}}
       
\author{Xu Feng\thanks{Current address:\ KEK.}\\
NIC, DESY, Platanenallee 6, D-15738 Zeuthen, Germany\\
Universit\"at M\"unster,\hspace{-1pt} Institut f\"ur Theoretische Physik,\hspace{-1pt} Wilhelm-Klemm-Strasse 9, D-48149, Germany}

\author{Karl Jansen\\
NIC, DESY, Platanenallee 6, D-15738 Zeuthen, Germany}

\author{Marcus Petschlies\\
Institut f\"ur Physik, Humboldt-Universitaet zu Berlin, D-12489, Berlin, Germany}

\abstract{We calculate the leading-order hadronic correction to the
  anomalous magnetic moments of each of the three charged leptons in
  the Standard Model:\ the electron, muon and tau.  Working in
  two-flavor lattice QCD, we address essentially all sources of
  systematic error:\ lattice artifacts, finite-size effects,
  quark-mass extrapolation, momentum extrapolation and disconnected
  diagrams.  The most significant remaining systematic error, the exclusion of the
  strange and charm quark contributions, will be addressed in our
  four-flavor calculation.  We achieve a statistical accuracy of $2\%$ or better
  for the physical values for each of the three leptons and
  the systematic errors are at most comparable.
\vspace{0pt}
\begin{center}
\includegraphics[width=100pt]{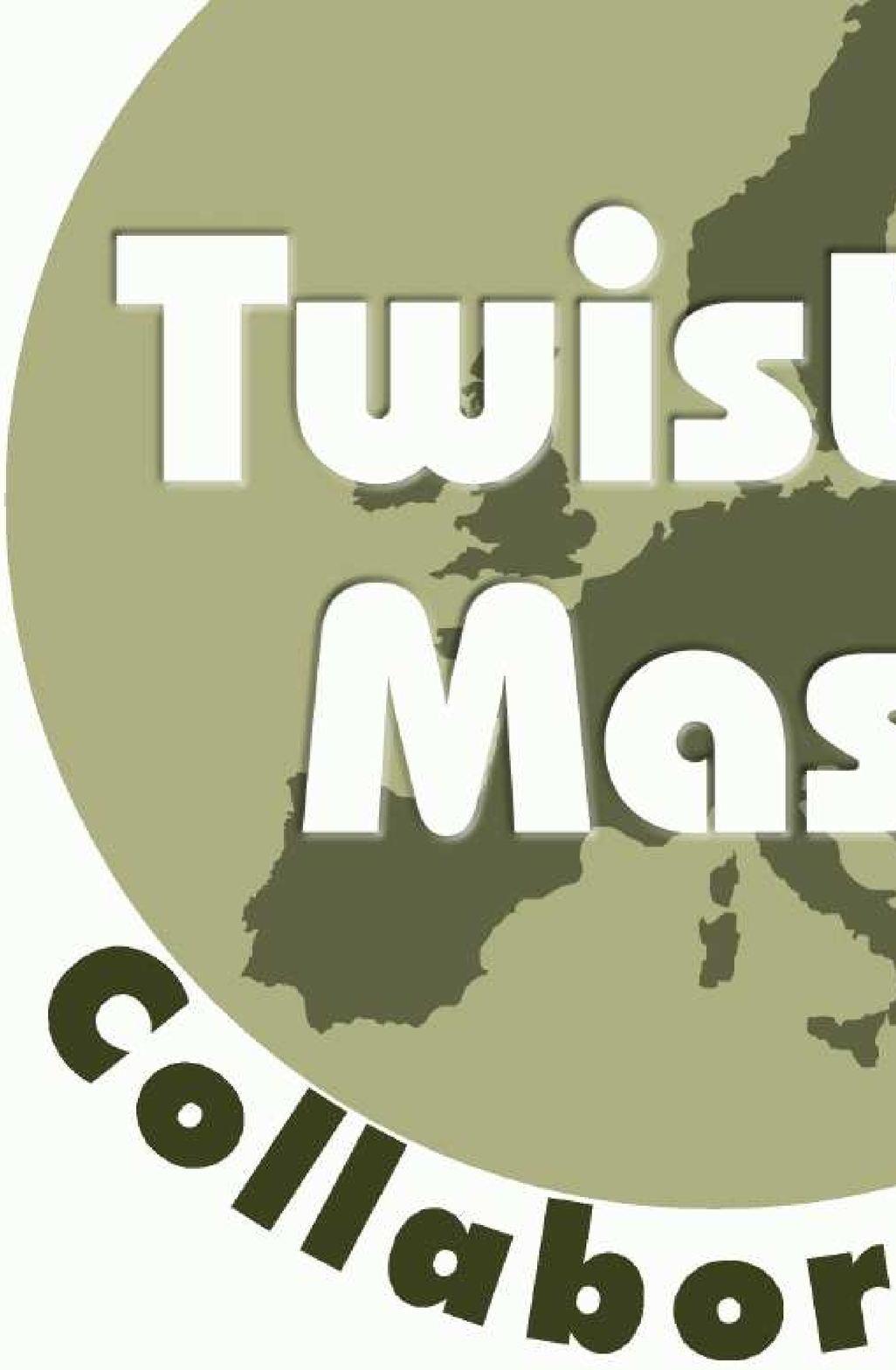}
\end{center}
}

\FullConference{35th International Conference of High Energy Physics - ICHEP2010,\\
July 22-28, 2010\\
Paris France}

\begin{document}

\section{Introduction}

The lingering $3\sigma$ discrepancy between the
measured~\cite{Bennett:2006fi} and
calculated~\cite{Jegerlehner:2009ry} values of the anomalous magnetic
moment of the muon $\amu$ raises the possibility of physics beyond the
Standard Model.  However, a clear understanding of the significance of
this discrepancy is complicated by the fact that the Standard Model
calculation uses additional experimental inputs and models to account
for the hadronic contributions to $\amu$.
Additionally, the hadronic corrections give rise to the
dominant sources of error in the Standard Model computation.  Thus a
non-perturbative first-principles calculation of the hadronic
contributions to $\amu$ could clarify the significance of the
current $3\sigma$ effect and potentially reduce the error of the
Standard Model result.

The hadronic contributions to $\amu$ are organized as an expansion in
the electromagnetic coupling $\alpha$.  The leading-order contribution
is responsible for the largest single source of error in the Standard
Model prediction for $\amu$.  In this work, we calculate the
leading-order hadronic correction to $\amu$ using two-flavor lattice
QCD.  We use the gauge field ensembles of the European Twisted Mass Collaboration~\cite{Baron:2009wt}.  
By introducing a new method to overcome the difficulties of
previous calculations, we achieve a statistical precision of less than $2\%$.
All sources of systematic error have been examined and appear to be no
larger than the statistical error.  As a check, we
calculate the leading-order correction for the electron and tau
leptons and reach similar conclusions.  The final results of our two-flavor
calculation, including an estimate of the systematic errors, will be
presented in~\cite{article}, so here we summarize only the
recent advances in our calculation.

\section{Leading-order hadronic contribution}

The expression for the
leading-order correction to the anomalous magnetic moment $\alhvp$ of
a lepton $l$ with mass $m_l$ was given by Blum~\cite{Blum:2002ii} as
\be
\label{alhvp}
\alhvp = \alpha^2\!\! \int_0^{\infty}\!\!\!\! dQ^2\, \frac{1}{Q^2} w(Q^2/m_l^2)\, \Pi_R(Q^2)
\ee
where $\Pi_R(Q^2) = \Pi(Q^2) - \Pi(0)$ is the renormalized vacuum
polarization $\Pi(Q^2)$, $Q^2$ is the Euclidean momentum and $w(Q^2/m_l^2)$ is a known function.  The
computation of $\Pi(Q^2)$ is now a standard calculation in lattice
QCD.  The details of our approach were recently described
in~\cite{Renner:2010zj}.

\section{Extrapolation and interpolation}

\begin{figure}
\begin{minipage}{210pt}
\includegraphics[width=210pt]{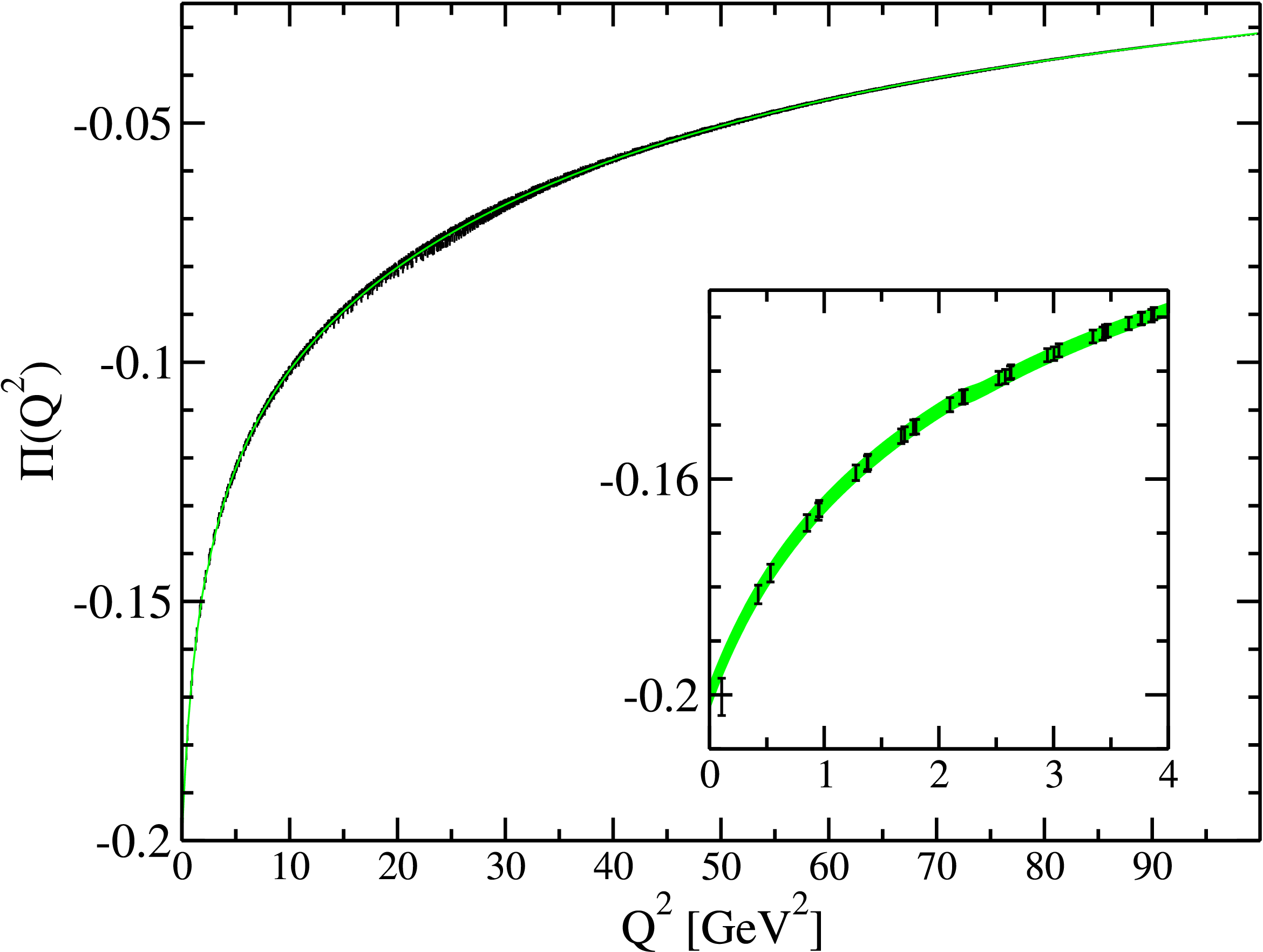}\vspace{5pt}
\caption{Vacuum polarization function, $\Pi(Q^2)$.  This is an example
  of our calculation of $\Pi(Q^2)$ and the interpolation of the entire range of $Q^2$ that is calculated on the lattice.}
\label{pi}
\end{minipage}
\hspace{4pt}
\begin{minipage}{210pt}
\includegraphics[width=210pt]{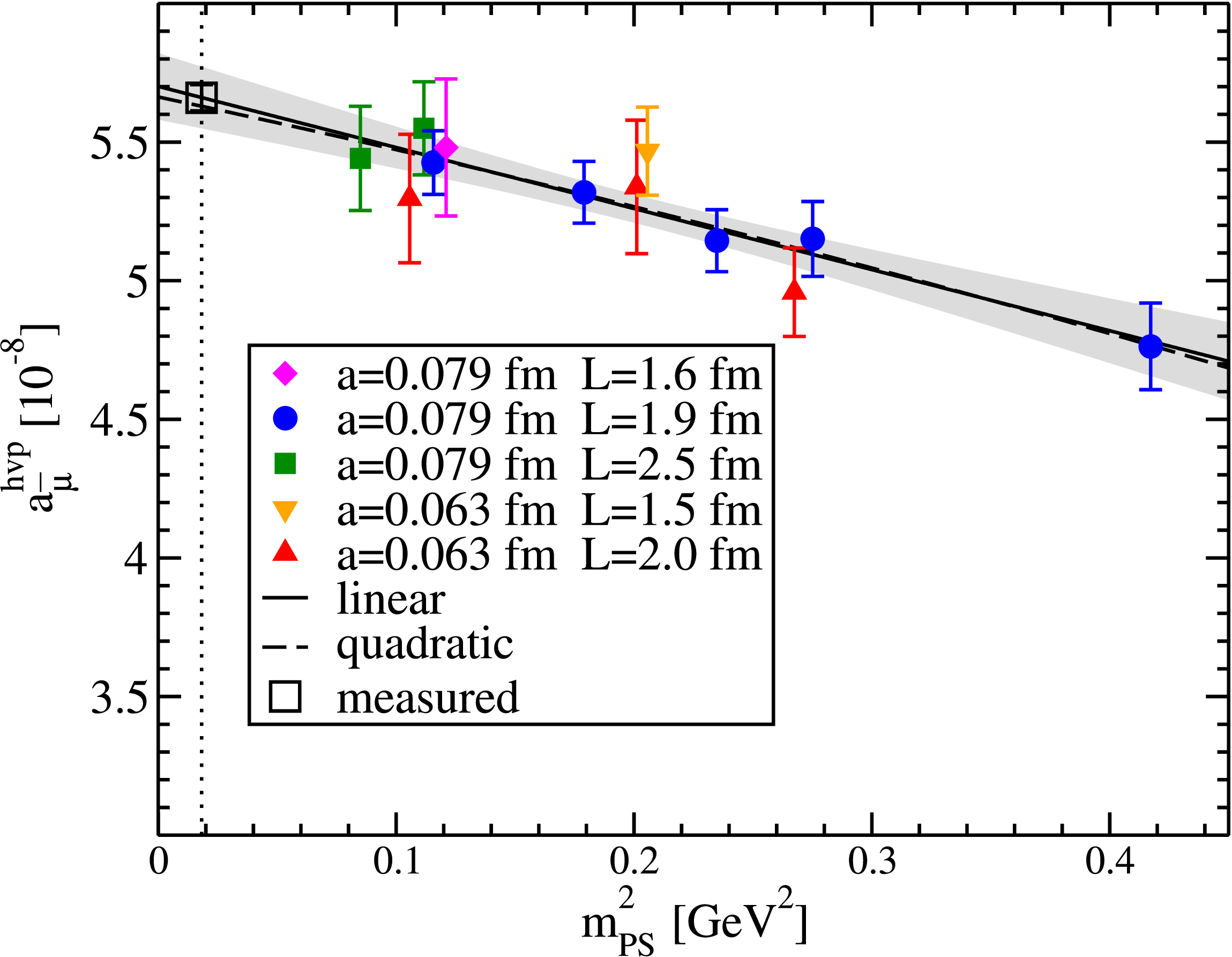}
\caption{Leading-order hadronic correction for the muon, $\amuhvp$.
  This is our calculation of $\amuhvp$, using the modified method
  described in the text, compared to the two-flavor contribution to
  the experimental value.}
\label{amubar}
\end{minipage}
\end{figure}

The lattice calculation of $\Pi(Q^2)$ is restricted to discrete values of $Q^2$ and must be interpolated and
extrapolated to $Q^2=0$ to perform the integration in
\Eq{\ref{alhvp}}.  To describe the high $Q^2$ region of $\Pi(Q^2)$, we
use
\bd
\Pi_\mathrm{high}(Q^2) = c + \ln Q^2 \sum_{n=0}^{O} b_n (Q^2)^n\,.
\ed
The constant $c$ is necessary because $\Pi(Q^2)$ by itself is
ultraviolet divergent.  The expected $\ln Q^2$ behavior is explicitly
accounted for, but the polynomial sum is added to ensure that
$\Pi_\mathrm{high}$ provides a complete basis in which to expand.  The
low $Q^2$ region of $\Pi(Q^2)$ is dominated by the contributions from
the lightest vector-meson states.  Therefore, we take the simple
tree-level form used in effective field theories of
vector-mesons~\cite{Aubin:2006xv} and then add a polynomial expansion
to provide a complete basis of functions.  The resulting expression is
\bd
\label{pilow}
\Pi_{\mathrm{low}}(Q^2) = -\frac{5}{9} \sum_{i=1}^{M} g_{Vi}^2 \frac{m_{Vi}^2}{Q^2 + m_{Vi}^2} + \sum_{n=0}^{N} a_n (Q^2)^n\,.
\ed
The sum over $i$ gives the tree-level contribution for $M$ vector
mesons with masses $m_{Vi}$ and couplings $g_{Vi}$.  These masses and
couplings have precise meanings of their own and are calculated in the
same lattice computation.  The form for $\Pi_\mathrm{low}$ is then
matched to $\Pi_\mathrm{high}$ to provide a complete description
that is suitable for numerical integration.  An example of this interpolation
and extrapolation is shown in \Fig{\ref{pi}}.  This method gives a fully
non-perturbative determination of $\alhvp$ that does not rely on QCD perturbation theory.

\section{Modified method}

\begin{figure}
\begin{minipage}{210pt}
\includegraphics[width=210pt]{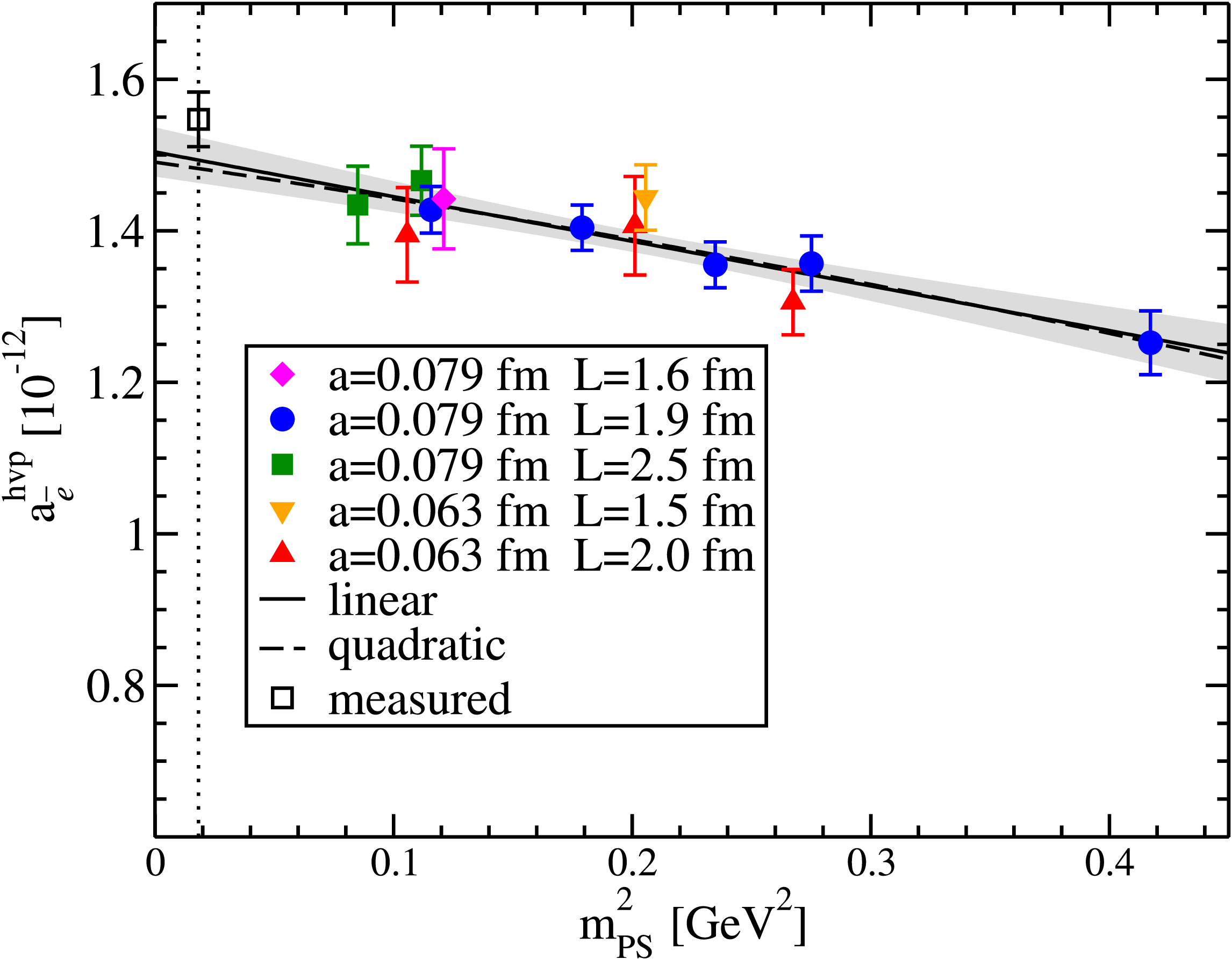}
\caption{Leading-order hadronic correction for the electron, $\aehvp$.
  This is the same as in \Fig{\protect\ref{amubar}} but now for the
  electron.}
\label{aebar}
\end{minipage}
\hspace{4pt}
\begin{minipage}{210pt}
\includegraphics[width=210pt]{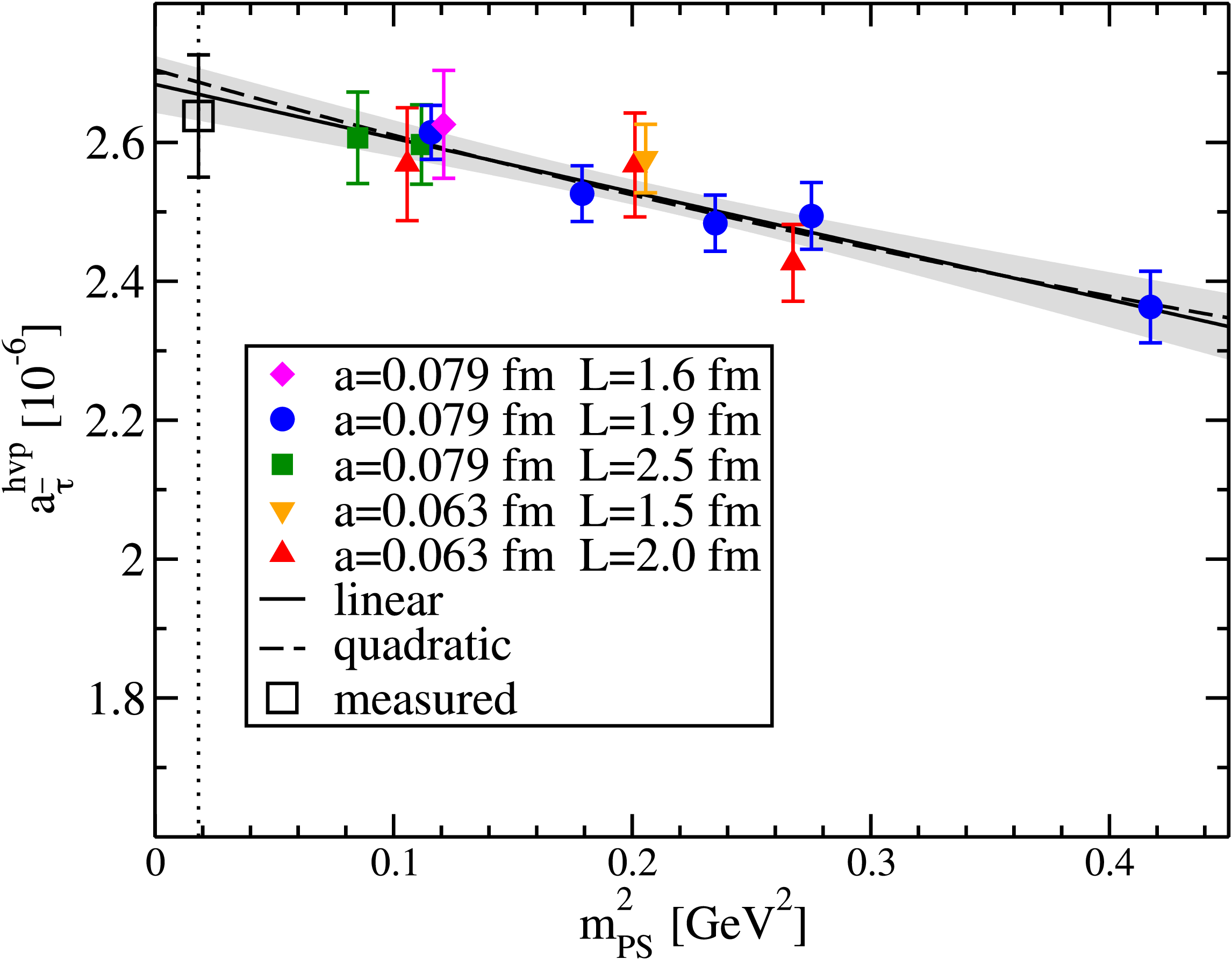}
\caption{Leading-order hadronic correction for the tau, $\atauhvp$.
  This is the same as in \Fig{\protect\ref{amubar}} but now for the
  tau.}
\label{ataubar}
\end{minipage}
\end{figure}

We introduce a class of quantities that have the same physical limit
as $\alhvp$ but approach that limit more smoothly as a function of the pseudo-scalar meson mass $m_{PS}$.
For
any hadronic quantity $H$, we define
\bd
\albarhvp = \alpha^2\!\! \int_0^{\infty}\!\!\!\! dQ^2\, \frac{1}{Q^2} w(Q^2/m_l^2\cdot H_\mathrm{phys}^2 / H^2)\, \Pi_R(Q^2)
\ed
where $H$ is understood to be calculated at each value of $m_{PS}$ and
$H_\mathrm{phys}= H(m_{PS}\rightarrow m_\pi)$.  By construction
$\albarhvp = \alhvp$ as $m_{PS}$ approaches $m_\pi$.  Any choice of $H$ leads to a
valid definition but $H=m_V$, the lightest vector-meson mass, leads to
a mild $m_{PS}$ dependence and results in a well-controlled
lattice calculation of $\alhvp$ in the physical limit.  The results
for $\amubarhvp$ are shown in \Fig{\ref{amubar}}.  The two-flavor
contribution to the experimentally measured value of $\amu$~\cite{Jegerlehner:2009ry,article} is also
shown in the plot and we find good agreement between the linearly
extrapolated lattice calculation and the measured value.  As a check
of the method, we perform the same calculation and comparison for the
electron in \Fig{\ref{aebar}} and for the tau in \Fig{\ref{ataubar}}.

\section{Conclusions}

We have calculated the two-flavor contribution to the leading-order
hadronic correction to the magnetic moments of the three charged
leptons in the Standard Model:\ the electron, muon and tau.  Combining
an extensive study of all the systematics with a modification to the
previously used method, we can reliably reproduce the two-flavor contribution to the experimentally measured values.
The next systematic error that needs to be addressed, the omission of the strange and
charm quark contributions, will be resolved by our ongoing four-flavor lattice
calculation.  This will open the possibility of a completely
first-principles determination of the leading-order hadronic contribution to
the Standard Model values for the lepton magnetic moments.

\begin{acknowledgments}
We are grateful to the John von Neumann Institute for Computing (NIC),
the J{\"u}lich Supercomputing Center and the DESY Zeuthen Computing
Center for their computing resources and support.  This work has been
supported in part by the DFG
Sonder\-for\-schungs\-be\-reich/Transregio SFB/TR9-03, the DFG project
Mu 757/13 and the U.S. DOE under Contract No. DE-AC05-06OR23177.
\end{acknowledgments}

\bibliography{gm2}

\begin{thebibliography}{1}

\bibitem{Bennett:2006fi}
Muon G-2 Collaboration, G.~Bennett {\em et~al.},
\newblock Phys. Rev. {\bf D73}, 072003 (2006), hep-ex/0602035.

\bibitem{Jegerlehner:2009ry}
F.~Jegerlehner and A.~Nyffeler,
\newblock Phys. Rept. {\bf 477}, 1 (2009), arXiv:0902.3360.

\bibitem{Baron:2009wt}
ETMC Collaboration, R.~Baron {\em et~al.},
\newblock JHEP {\bf 08}, 097 (2010), arXiv:0911.5061.

\bibitem{article}
X.~Feng, K.~Jansen, M.~Petschlies, and D.~B. Renner,
\newblock in preparation.

\bibitem{Blum:2002ii}
T.~Blum,
\newblock Phys. Rev. Lett. {\bf 91}, 052001 (2003), hep-lat/0212018.

\bibitem{Renner:2010zj}
D.~B. Renner, X.~Feng, K.~Jansen, and M.~Petschlies,
\newblock PoS {\bf LAT2010}, 155 (2010), arXiv:1011.4231.

\bibitem{Aubin:2006xv}
C.~Aubin and T.~Blum,
\newblock Phys. Rev. {\bf D75}, 114502 (2007), hep-lat/0608011.

\end{thebibliography}
\bibliographystyle{h-physrev}

\end{document}